\documentclass[aps,twocolumn,showpacs,prb,reprint,amsmath,longbibliography,superscriptaddress]{revtex4-2}
            
\usepackage{bbm}
\usepackage{mathrsfs}
\usepackage{amsmath}
\usepackage{amsfonts}
\usepackage[colorlinks=true, 
            citecolor=blue, 
            linkcolor=blue, 
            urlcolor=blue]{hyperref}
\usepackage{graphicx,epstopdf}
\usepackage{subfigure}
\usepackage{epsfig}
\usepackage{dcolumn}
\usepackage{bm}
\usepackage{color}
\usepackage{natbib}
\usepackage{amssymb}  
\usepackage{xcolor}
\usepackage{braket}
     
\begin{document}

\title{Failure of Mean-Field Critical Scaling and
Two Distinct Dissipation Regimes in La$_2$SmNi$_2$O$_{7-\delta}$}

\author{Yongze Ye}
\affiliation{
Center for Superconducting Physics and Materials,
National Laboratory of Solid State Microstructures
and Department of Physics,
Collaborative Innovation Center for Advanced Microstructures,
Nanjing University, Nanjing 210093, China
}

\author{Feiyu Li}
\affiliation{
State Key Laboratory of Crystal Materials,
Institute of Crystal Materials,
Shandong University, Jinan, China
}

\author{Di Peng}
\affiliation{
Center for High Pressure Science and Technology Advanced Research, Shanghai, China
}
\affiliation{
Shanghai Key Laboratory of Material Frontiers Research in Extreme Environments (MFree),
Shanghai Advanced Research in Physical Sciences (SHARPS),
Shanghai, China
}

\author{Junjie Zhang}
\email{junjie@sdu.edu.cn}
\affiliation{
State Key Laboratory of Crystal Materials,
Institute of Crystal Materials,
Shandong University, Jinan, China
}

\author{Hai-Hu Wen}
\email{hhwen@nju.edu.cn}
\affiliation{
Center for Superconducting Physics and Materials,
National Laboratory of Solid State Microstructures and Department of Physics,
Collaborative Innovation Center for Advanced Microstructures,
Nanjing University, Nanjing 210093, China
}
\date{\today}

\begin{abstract}
Recently, superconductivity with an onset transition temperature ($T_{\mathrm{c}}^{\mathrm{onset}}$) up to 96 K has been observed in the $\mathrm{La}_{2-x}\mathrm{Sm}_{1+x}\mathrm{Ni}_2\mathrm{O}_{7-\delta}$ system under pressure, providing an attractive platform for investigating superconducting mechanisms in nickelates.
Here, we systematically analyzed the critical scaling and vortex dynamics for the $\mathrm{La}_2\mathrm{Sm}\mathrm{Ni}_2\mathrm{O}_{7-\delta}$ single crystal with $T_{\mathrm{c}}^{\mathrm{onset}}=92.5$ K under pressure.
Within the Ullah--Dorsey framework, neither the two-dimensional nor three-dimensional scaling provides a good collapse near $T_{\mathrm{c}}^{\mathrm{MF}}$.
This failure may be associated with the multiband composite order parameter of bilayer nickelates, in which pairing is mainly driven by antiferromagnetic exchange in the $d_{z^2}$ orbital, whereas the superfluid stiffness is mainly governed by the $d_{x^2-y^2}$ orbital.
The two-step decrease in the logarithmic resistivity reveals two distinct dissipative regimes. 
The low-temperature regime is well described by the thermally assisted flux flow (TAFF) model, whereas the high-temperature regime cannot be explained by the vortex dynamics involved with the superconducting fluctuation. The crossover between the two regimes may be associated with the orbital-dependent evolution of vortex cores.
These results suggest the significant role of the composite order parameter in critical fluctuations and vortex-motion dissipation in nickelates.

\end{abstract}

\maketitle
\section{Introduction}
Since the discovery of superconductivity in the infinite-layer nickelate thin film Nd$_{1-x}$Sr$_x$NiO$_2$ in 2019, nickelate superconductors have attracted considerable attention as a new class of unconventional superconductors~\cite{2019LiNature,2022GuInnovation}. 
In 2023, superconductivity with a transition temperature $T_{\rm c}$ approaching 80~K was reported in the bilayer nickelate La$_3$Ni$_2$O$_7$ under pressure above 14~GPa~\cite{2023SunNature}. 
This breakthrough extended nickelate superconductivity from infinite layer thin films to bulk single crystals of the Ruddlesden--Popper (RP) phase and brought $T_{\rm c}$ into the liquid-nitrogen range, thereby stimulating extensive experimental and theoretical studies of bilayer RP nickelates~\cite{2024WangCPL,2026PuphalNRP}.
In contrast to infinite-layer nickelates and cuprates, Ni in bilayer nickelates has an average valence of $+2.5$, corresponding to a $3d^{7.5}$ configuration~\cite{2023ZhangPRB}. 
Theoretical studies have shown that the bilayer NiO$_2$ planes lead to strong interlayer hybridization between the Ni-$3d_{z^2}$ orbitals mediated by the apical O-$2p_z$ orbitals, resulting in a bonding--antibonding splitting. 
Meanwhile, the more dispersive in-plane Ni-$3d_{x^2-y^2}$ bands are partially occupied near the Fermi level~\cite{2023LuoPRL}.
Therefore, the electronic states near the Fermi level are dominated by Ni-$3d_{z^2}$, Ni-$3d_{x^2-y^2}$, and O-$2p$ orbitals, which has motivated investigations of the respective roles of the two Ni orbitals in the superconducting pairing mechanism~\cite{2025HuCP,2025GuPRB,2024LuoNPJ,2023YangPRB}.

Electronic structure calculations under high pressure have predicted three principal Fermi surface pockets. The $\alpha$ and $\beta$ pockets primarily exhibit $d_{x^2-y^2}$ character with some weight from $3d_{z^2}$ orbitals near the (0,$\pi$) or ($\pi$,0) momenta on the $\beta$ pocket, whereas the $\gamma$ pocket has a dominant $3d_{z^2}$ contribution~\cite{2023LuoPRL,2023ZhangPRB}. 
These results reveal the multiband character of the bilayer RP nickelates, in contrast to the single band electronic structure of the cuprates. 
Experimental evidence at ambient pressure further supports the multiband picture.
ARPES measurements on the bilayer nickelate directly resolve the $\alpha$ electron pocket around the $\Gamma$ point and the $\beta$ hole pocket around the $M$ point, while the relatively flat $\gamma$ band with Ni-$3d_{z^2}$ character lies approximately 50~meV below the Fermi level~\cite{2024YangNC}.
Optical conductivity measurements further reveal two Drude components, consistent with a multiband electronic structure involving the Ni-$3d_{x^2-y^2}$ and Ni-$3d_{z^2}$ orbitals~\cite{2024LiuNC}.
More recently, scanning tunneling microscopy (STM) measurements on $\mathrm{La}_2\mathrm{Pr}\mathrm{Ni}_2\mathrm{O}_{7-\delta}$ thin films revealed two distinct superconducting gaps, consistent with the multiband scenario~\cite{2026FanSA}.

Recently, Sm substituted bilayer nickelate single crystals, (La,Sm)$_3$Ni$_2$O$_7$, were synthesized using an ambient pressure flux method, and pressure induced superconductivity above the liquid nitrogen temperature was observed~\cite{2026LiNature}. 
By tuning the Sm content, an onset transition temperature of 96~K was achieved in $\mathrm{La}_{1.57}\mathrm{Sm}_{1.43}\mathrm{Ni}_2\mathrm{O}_{7-\delta}$. These results establish Sm substituted bilayer nickelates as a promising system for investigating superconducting mechanisms in nickelates.
In this letter, we report anomalous transport behavior in $\mathrm{La}_{2}\mathrm{Sm}\mathrm{Ni}_2\mathrm{O}_{7-\delta}$ under pressure.
In the superconducting fluctuation regime, neither the two-dimensional nor the three-dimensional Ullah--Dorsey scaling produces a good data collapse. 
Introducing additional field-dependent power-law factors, however, leads to a significantly improved collapse in both 2D and 3D scaling.
In the regime governed by vortex dynamics, the apparent activation energy $U_{app}$ exhibits two distinct plateaus, consistent with the two-stage decrease observed in the logarithmic resistivity, indicating a crossover between different dissipation mechanisms upon cooling.
Our analysis suggests that the multiorbital nature of superconductivity produces rich transport properties in the $\mathrm{La}_{2}\mathrm{Sm}\mathrm{Ni}_2\mathrm{O}_{7-\delta}$ system, making it a promising platform for investigating unconventional high-temperature superconductivity.
\section{Experimental details}
La$_2$SmNi$_2$O$_{7-\delta}$ single crystals were synthesized at ambient pressure using a flux method. As-grown crystals were post-annealed at $p_{\mathrm{O}_2}=1.5~\mathrm{bar}$ for 10 days to optimize oxygen content. Sample 1, which was annealed under these conditions, was selected for the transport measurements presented in this work, the raw data have been published in Ref.~\cite{2026LiNature}. Further details of the crystal growth and annealing procedures are also provided there. Resistivity measurements under pressure were performed using a BeCu diamond anvil cell (DAC), with helium serving as the pressure transmitting medium (PTM). The resistivity as a function of temperature and magnetic field was measured using a Physical Property Measurement System (PPMS, Quantum Design) over the temperature range from 2 to 310~K and in magnetic fields up to 7~T.
\begin{figure}[t]
\centering
\hspace*{-0.3in}
\includegraphics[width=8cm]{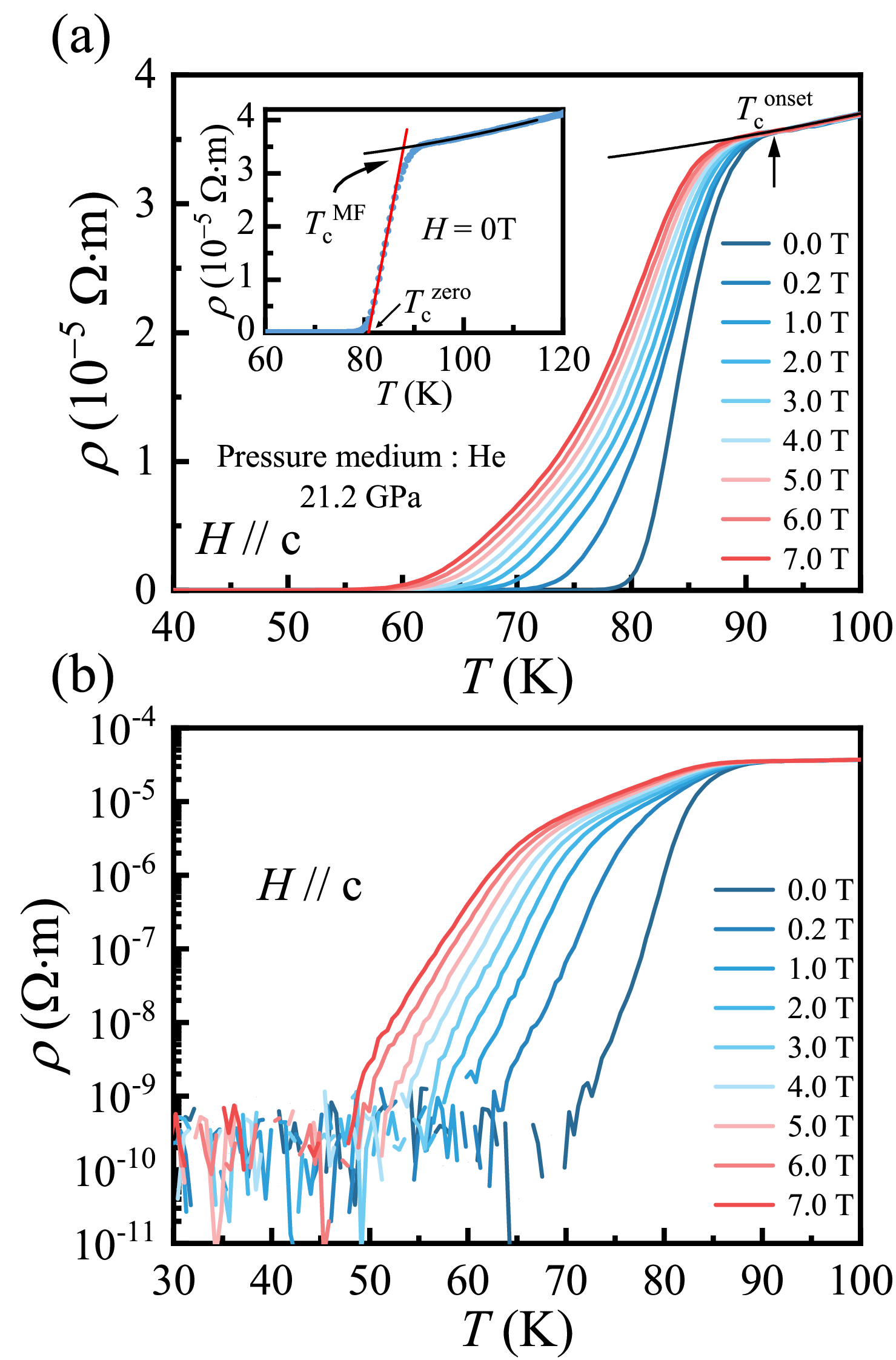}
\caption{Electrical resistivity of the La$_2$SmNi$_2$O$_{7-\delta}$ single crystal (Sample 1) as a function of temperature under magnetic fields up to 7 T with $H \parallel c$ plotted on linear scales (a) and semi-log scales (b), respectively. 
The inset of (a) shows the superconducting transition at zero magnetic field and the determination of the mean-field transition temperature $T_c^{\mathrm{MF}}$. The data is reproduced from ~\cite{2026LiNature}.
}
\label{fig:rhoT}
\end{figure}
\section{Results and discussion}
Figure~\ref{fig:rhoT}(a) presents the temperature dependent resistivity ($\rho$--$T$) of the $\mathrm{La_2SmNi_2O_{7-\delta}}$ single crystal under magnetic fields at $21.2~\mathrm{GPa}$. Curves obtained at different magnetic fields exhibit a nearly common crossing point, from which the onset transition temperature is determined as $T_c^{\mathrm{onset}}=92.5~\mathrm{K}$. The inset of Figure~\ref{fig:rhoT}(a) shows an enlarged view of the superconducting transition under zero magnetic field, with $T_c^{\mathrm{zero}}=82~\mathrm{K}$. 
For the critical scaling analysis presented below, the mean-field transition temperature $T_c^{\mathrm{MF}}=87.6~\mathrm{K}$ is defined as the intersection of the extrapolated linear fits from the normal-state and transition regions, which coincides with the value obtained using the $90\%\ \rho_n$ criterion.
With increasing magnetic field, the superconducting transition becomes significantly broadened. Such behavior arises from the combined effects of critical superconducting fluctuations near the transition\cite{1991IkedaJPSJ} and thermally activated flux motion\cite{1989NelsonPRB,1990VinokurPRL,1991FisherPRB,1991TesanovicPRB}. 
In the semi-logarithmic plot shown in Figure~\ref{fig:rhoT}(b), the resistivity displays a pronounced two-step decrease upon cooling, suggesting that the dissipation involves two distinct dynamical regimes. 
To elucidate the origins of these features, we systematically investigate the critical fluctuation behavior and vortex dynamics of the $\mathrm{La_2SmNi_2O_{7-\delta}}$ single crystal in the following sections.

\subsection*{A. Scaling of superconducting fluctuations}
\begin{figure*}[ht]
\centering
\hspace*{-0.2in}
\includegraphics[width=16cm]{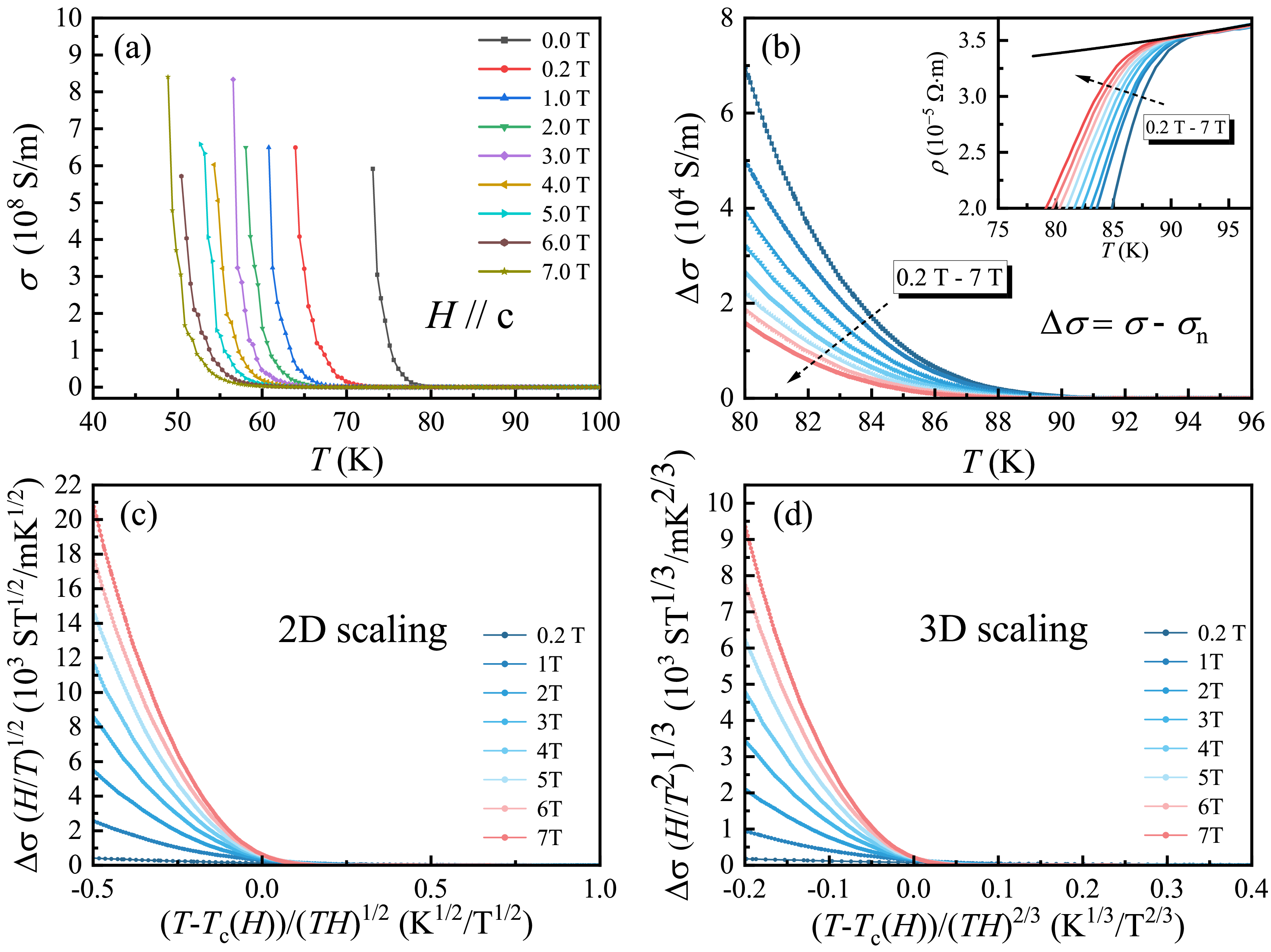}
\caption{
(a) Temperature dependence of electrical conductivity $\sigma$ under different magnetic fields. 
(b) Temperature dependence of excess conductivity $\Delta\sigma=\sigma-\sigma_n$. The inset shows the resistivity in the superconducting fluctuation regime, where the black solid line represents the normal-state background. (c) 2D and (d) 3D Ullah--Dorsey scaling of the excess conductivity. All measurements were performed on the $\mathrm{La_2SmNi_2O_{7-\delta}}$ single crystal at 21.2 GPa, with $H \parallel c$.
}
\label{fig:scaling}
\end{figure*}
According to the theory of superconducting fluctuations, transient Cooper pairs are generated by thermal fluctuations above $T_{\rm c}$ and contribute an excess conductivity in an applied electric field. In conventional superconductors, such effects are difficult to observe because the thermal energy $k_BT_c$ is much smaller than the condensation energy $H_c(0)^2\xi^3$ in a coherence volume, here $H_c(0)$ is the thermodynamic field, $\xi$ is the coherence length\cite{1968AslamasovPLA}. However, in high-$T_{\rm c}$ superconductors, these effects are greatly enhanced by the extremely short coherence length, strong anisotropy, and high $T_{\rm c}$, making the experimental observation possible~\cite{1990WenAPS,2000NaughtonPRB}.

Ullah and Dorsey developed a universal scaling framework for the superconducting fluctuation under magnetic fields based on the time-dependent Ginzburg--Landau (TDGL) theory and a self-consistent Hartree approximation \cite{1990UllahPRL,1991UllahPRB}. 
In this framework, the excess conductivity for two-dimensional (2D) and three-dimensional (3D) superconducting fluctuations is expected to follow
\begin{equation}
\Delta\sigma_{\rm 2D}(H)
=
\left(\frac{T}{H}\right)^{1/2}
F_{\rm 2D}
\left[
A\frac{T-T_{\rm c}(H)}{(TH)^{1/2}}
\right],
\label{eq:UD2D}
\end{equation}
and
\begin{equation}
\Delta\sigma_{\rm 3D}(H)
=
\left(\frac{T^2}{H}\right)^{1/3}
F_{\rm 3D}
\left[
B\frac{T-T_{\rm c}(H)}{(TH)^{2/3}}
\right],
\label{eq:UD3D}
\end{equation}
respectively. Here, $F_{\rm 2D}$ and $F_{\rm 3D}$ are universal scaling functions, whereas $A$ and $B$ are coefficients related to the transition width of the material. 
This approach predicts that excess conductivity curves measured under different magnetic fields should collapse onto a single curve after appropriate scaling, which has successfully described the superconducting fluctuation of a number of superconductors \cite{1991WelpPRL,2012MarraNJP,2017AhmadSST,2018AhmadSR}.

Figure~\ref{fig:scaling}(a) shows the temperature dependence of the electrical conductivity $\sigma$ under different magnetic fields. 
To investigate the superconducting fluctuation behavior, we calculated the excess conductivity, defined as
\begin{equation}
\Delta\sigma(T,H)=\sigma(T,H)-\sigma_{\rm n}(T,H),
\label{eq:excess}
\end{equation}
where $\sigma$ is the measured conductivity and $\sigma_{\rm n}=1/\rho_{\rm n}$ is the conductivity of the normal state, with $\rho_{\rm n}$ obtained by extrapolating the normal-state resistivity, as indicated by the black solid line in the inset of Figure~\ref{fig:scaling}(b).
The resulting excess conductivity, shown in Figure~\ref{fig:scaling}(b), increases rapidly upon cooling and is suppressed with increasing magnetic field.
To examine the dimensionality of the superconducting fluctuations in the La$_2$SmNi$_2$O$_{7-\delta}$ single crystal, we analyzed the data using both 2D and 3D Ullah--Dorsey scaling.
\begin{figure}[t]
\centering
\hspace*{-0.5in}
\includegraphics[width=8cm]{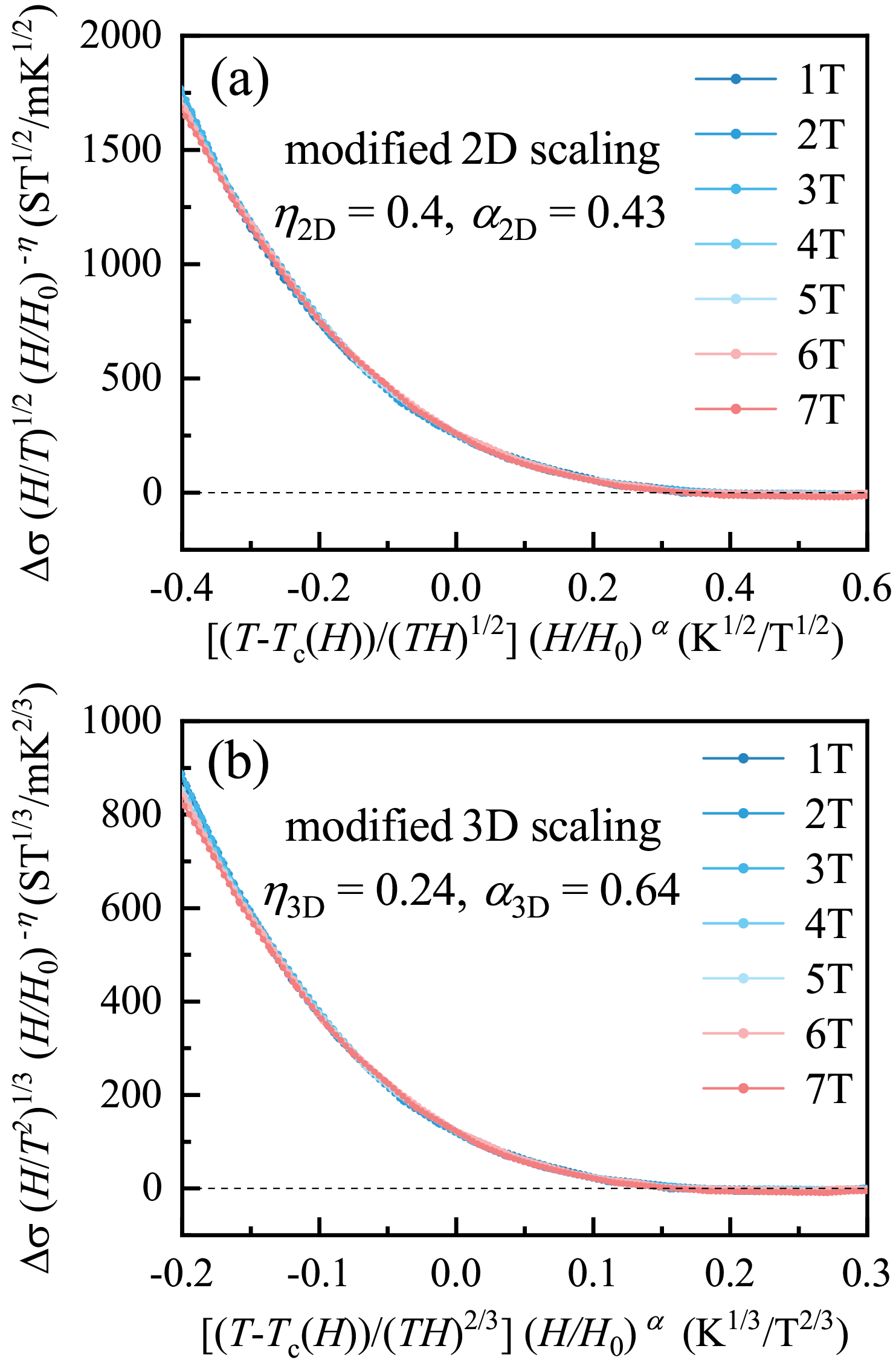}
\caption{
Optimized data collapse obtained using modified Ullah--Dorsey scaling of the excess conductivity.
(a) 2D scaling with $\alpha_{\mathrm{2D}}=0.43$ and $\eta_{\mathrm{2D}}=0.40$.
(b) 3D scaling with $\alpha_{\mathrm{3D}}=0.64$ and $\eta_{\mathrm{3D}}=0.24$.
The magnetic fields range from 1 to 7 T with $H\parallel c$, and $H_0=1$ T is the reference field.
}
\label{fig:modified 3D scale}
\end{figure}
Figures~\ref{fig:scaling}(c) and (d) present the 2D and 3D scaling results, respectively. However, neither of them produces a satisfactory collapse of the curves measured at different magnetic fields. The deviation is particularly pronounced on the lower temperature side of the scaling curves and for the lower magnetic fields. \par
The standard Ullah--Dorsey theory is based on a single order parameter $\Psi(\mathbf r)=|\Psi(\mathbf r)|e^{i\theta(\mathbf r)}$, with one effective mass and one characteristic coherence length. In this scenario, the pairing amplitude and the superfluid stiffness originate from the same electronic component.
Therefore, the failure of the scaling may be attributed to the composite order parameter of multibands in the system.
For bilayer nickelates, theoretical studies suggest that the pairing interaction and superfluid response may be governed by different orbitals. 
The pairing interaction may be associated primarily with the localized $d_{z^2}$ states, whereas the phase stiffness is mainly carried out by the more itinerant $d_{x^2-y^2}$ states~\cite{2023YangPRB,2025ZhengPRB}. These two orbitals are coupled by hybridization but may have different effective masses, coherence lengths, and field responses. Their contributions cannot simply be reduced to a single effective mass or a single fluctuation width, which naturally explains the failure of the scaling.
Furthermore, the breakdown of the lowest-Landau-level (LLL) approximation at low magnetic fields may provide an additional contribution to the deviations. In the Ullah--Dorsey framework, the characteristic fluctuation width is determined by $(TH)^{1/2}$ in 2D or $(TH)^{2/3}$ in 3D, which is valid only when the LLL condition is satisfied. However, higher Landau levels cannot be neglected at low fields, so the LLL scaling variable is no longer sufficient.

To further investigate the deviation from conventional Ullah--Dorsey scaling, we introduce additional dimensionless factors $(H/H_0)^{\alpha_d}$ and $(H/H_0)^{-\eta_d}$ into the horizontal and vertical scaling variables, respectively, where $d$ denotes the fluctuation dimensionality (2D or 3D) and $H_0=1$ T is a fixed reference field introduced to ensure dimensional consistency. The modified 2D scaling variables can be expressed as
\begin{equation}
X'_{\mathrm{2D}} =
\frac{T-T_{\mathrm{c}}(H)}{(TH)^{1/2}}
\left(\frac{H}{H_0}\right)^{\alpha_{\mathrm{2D}}},
\label{eq:modified_X_2D}
\end{equation}
and
\begin{equation}
Y'_{\mathrm{2D}} =
\Delta\sigma(T,H)
\left(\frac{H}{T}\right)^{1/2}
\left(\frac{H}{H_0}\right)^{-\eta_{\mathrm{2D}}}.
\label{eq:modified_Y_2D}
\end{equation}
Similarly, the modified 3D scaling variables are given by
\begin{equation}
X'_{\mathrm{3D}} =
\frac{T-T_{\mathrm{c}}(H)}{(TH)^{2/3}}
\left(\frac{H}{H_0}\right)^{\alpha_{\mathrm{3D}}},
\label{eq:modified_X_3D}
\end{equation}
and
\begin{equation}
Y'_{\mathrm{3D}} =
\Delta\sigma(T,H)
\left(\frac{H}{T^2}\right)^{1/3}
\left(\frac{H}{H_0}\right)^{-\eta_{\mathrm{3D}}}.
\label{eq:modified_Y_3D}
\end{equation}

Figure~\ref{fig:modified 3D scale}(a) shows that the best collapse within the 2D framework is obtained with $\alpha=0.43$ and $\eta=0.40$, while Figure~\ref{fig:modified 3D scale}(b) shows that the optimized 3D scaling is achieved with $\alpha=0.64$ and $\eta=0.24$.
Similar scaling results for sample 2 are presented in the Supplemental Material~\cite{SM}.
For the modified 2D scaling, the field dependence in the horizontal scaling variable is reduced from $(H/H_0)^{-1/2}$ to approximately $(H/H_0)^{-0.07}$, while that in the vertical scaling variable is weakened from $(H/H_0)^{1/2}$ to $(H/H_0)^{0.10}$. Similarly, in the 3D framework, the corresponding field dependencies are reduced from $(H/H_0)^{-2/3}$ to $(H/H_0)^{-0.03}$ and from $(H/H_0)^{1/3}$ to $(H/H_0)^{0.09}$, respectively. 
In both cases, the weak residual field dependence of the modified horizontal scaling variable suggests that the magnetic field primarily shifts $T_c(H)$ toward lower temperatures, with little effect on the characteristic temperature region for the superconducting fluctuation. Meanwhile, the reduced field dependence of the vertical scaling variable suggests a much weaker suppression of the excess conductivity than predicted by conventional Ullah--Dorsey scaling.

\begin{figure*}[t]
\centering
\hspace*{-0.1in}
\includegraphics[width=16cm]{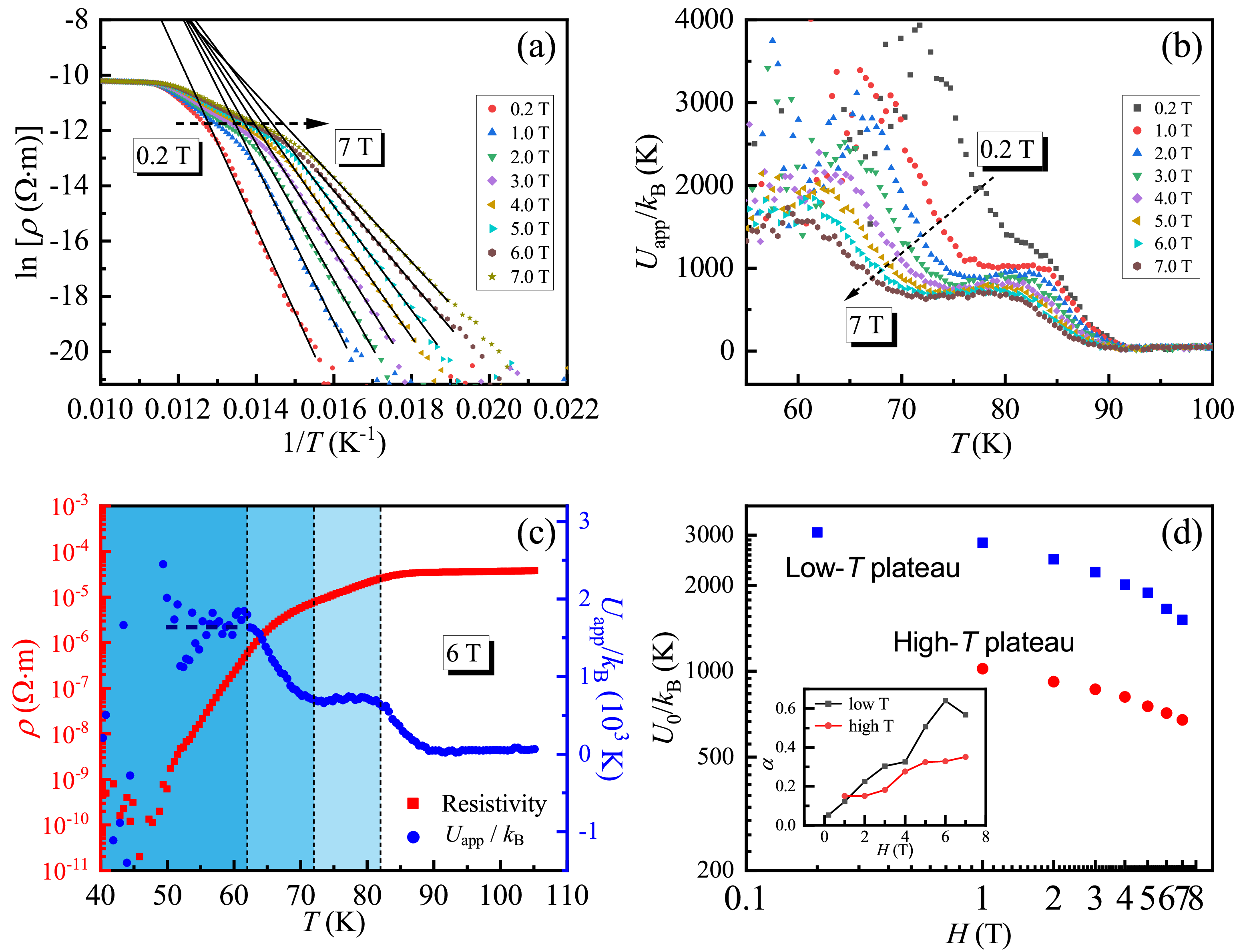}
\caption{
(a) Arrhenius plots of the resistivity of the La$_2$SmNi$_2$O$_{7-\delta}$ single crystal under magnetic fields up to 7 T. The black solid lines represent linear fits to the low-temperature regions.
(b) Temperature dependence of the apparent activation energy $U_{\mathrm{app}}(T,H)/k_{\mathrm B}$ under magnetic fields ranging from 0.2 to 7 T.
(c) Temperature dependence of $U_{\mathrm{app}}(T,H)/k_{\mathrm B}$ and the resistivity on a logarithmic scale at $H=6$ T. The blue dashed line is a guide to the eye.
(d) Magnetic-field dependence of the effective activation energies $U_0(H)/k_{\mathrm B}$ for the low- and high-temperature plateaus, determined by averaging $U_{\mathrm{app}}(T,H)/k_{\mathrm B}$ over the corresponding plateau regions in (b). The inset shows the field dependence of the local exponent $\alpha=-d\ln U_0/d\ln H$ for the two regimes.
}
\label{fig:TAFF}
\end{figure*}

The comparable quality of the modified 2D and 3D scaling collapses can be understood from the  reduction of the field dependence in both frameworks. Although the conventional 2D and 3D scaling relations involve different field exponents, the corrections nearly compensate for these differences, resulting in similar residual field dependence. The remaining difference in their temperature dependence is rather small over the narrow fluctuation regime.

This anomalous behavior may be attributed to the multi-orbital nature of the composite superconducting order parameter.
The relatively robust pairing correlations associated with the $d_{z^2}$ orbital may stabilize the superconducting fluctuation regime, whereas the $d_{x^2-y^2}$ orbital, which contributes significantly to the in-plane superfluid stiffness, is more susceptible to field-induced suppression of phase coherence. Consequently, the magnetic field can considerably reduce $T_c(H)$ while producing only weak changes in the fluctuation width and conductivity amplitude. This orbital-dependent response may explain the deviation from conventional Ullah--Dorsey scaling.

\subsection*{B. Thermally assisted flux flow  analysis}
The two-step resistive transition raises the question of whether the dissipation in the two temperature regimes can be described by the same thermally activated process. We therefore examine the resistivity using the thermally assisted flux flow (TAFF) model. For vortex motion over pinning barriers, the resistivity is expressed as
\begin{equation}
\rho(T,H)=\rho_{0f}
\exp\left[-\frac{U(T,H)}{k_{\rm B}T}\right],
\label{eq:TAFF}
\end{equation}
where $U(T,H)$ is the activation energy and $\rho_{0f}$ is a weakly temperature-dependent prefactor~\cite{1994BlatterRMP,1988PalstraPRL,1990PalstraPRB}.

Taking a linear temperature dependence of the activation barrier,
\begin{equation}
U(T,H)=U_0(H)\left(1-\frac{T}{T_c}\right),
\label{eq:Utemp}
\end{equation}
we obtain
\begin{equation}
\ln\rho(T,H)
=
\ln\rho_0(H)-\frac{U_0(H)}{k_{\rm B}T},
\label{eq:arrhenius}
\end{equation}
with $\ln\rho_0(H)=\ln\rho_{0f}+U_0(H)/(k_{\rm B}T_c)$.
For a single thermally activated process, Eq.~\eqref{eq:arrhenius} predicts a linear dependence of $\ln\rho$ on $1/T$, and the magnitude of the slope determines the effective activation energy $U_0(H)/k_{\mathrm B}$.

Figure~\ref{fig:TAFF}(a) shows the Arrhenius plot for La$_2$SmNi$_2$O$_{7-\delta}$, which reveals two linear intervals below the superconducting fluctuation regime. To display the evolution more directly, we calculate a temperature dependent apparent activation energy,
\begin{equation}
{U_{\rm {app}}(T,H)}/{k_{\rm B}}
=
-\frac{d\ln\rho}{d(1/T)}.
\label{eq:Ueff}
\end{equation}
As shown in Figure~\ref{fig:TAFF}(b), the $U_{app}/k_{\rm B}$--$T$ curves under each magnetic field except for $H = 0.2$ T reveal two plateau-like regimes separated by a broad crossover. 
Taking $H = 6$ T as an example, which is shown in Figure~\ref{fig:TAFF}(c), although the signal-to-noise ratio is reduced as the temperature approaches $T_c^{\mathrm{zero}}$, the two plateau-like regimes remain distinguishable. It can also be seen that the crossover between the two $U_{\mathrm{app}}$ plateaus occurs over the same temperature range as the gradual change in the slope of the logarithmic resistivity versus temperature.

Two TAFF regimes have previously been reported in a number of unconventional superconductors.
In YBa$_2$Cu$_3$O$_{7-\delta}$ films and YBa$_2$Cu$_4$O$_8$ crystals, different field dependence of the activation energy was associated with a crossover from a 3D vortex liquid to a decoupled quasi 2D vortex liquid~\cite{1995QiuPhysRevB,1998QiuPhysRevB}. 
In the RbCa$_2$Fe$_4$As$_4$F$_2$ system, the crossover between the two TAFF regimes is interpreted as a transition from plastic pinning in the high-temperature region to collective pinning dominated by planar defects in the low-temperature region~\cite{2020XingSST}.
For TAFF, $U_0(H)$ in a 3D vortex system is often described by $U_0(H)\propto H^{-\alpha}$, where the exponent $\alpha\sim0.5$--1~\cite{1992KuceraPRB,1993ChinPSC}. In a 2D or strongly layered system, $U_0(H) \propto \ln H$ is expected~\cite{1992JensenEPL,2002EngelPRB}.
However, the behavior observed here follows neither of these forms. As shown in Figure~\ref{fig:TAFF}(d), within each plateau-like regime, the $U_0(H)$ data exhibit downward curvature on double-logarithmic coordinates and therefore cannot be represented by a single relation of the form $U_0(H)\propto H^{-\alpha}$ or $U_0(H) \propto \ln H$, suggesting that the field dependence of the activation energy in this system deviates from both the 2D and 3D predictions.

To analyze the field dependence, we use the 3D form $U_0(H)\propto H^{-\alpha}$ as a reference and determine the local exponent $\alpha=-d\ln U_0/d\ln H$.
As shown in the inset of Figure~\ref{fig:TAFF}(d), in the low-temperature region, the exponent $\alpha_{\mathrm{low}}$ increases approximately from 0.05 to 0.64 with increasing field. In the high-temperature regime, however, $\alpha_{\mathrm{high}}$ increases only from 0.15 to 0.35 with increasing magnetic field, significantly smaller than the values expected for collective vortex creep ($\alpha \approx 1$) and plastic vortex creep ($\alpha \approx 0.5$). This discrepancy suggests that the dissipation in the high-temperature regime cannot be explained by conventional vortex dynamics and may indicate a different dissipation mechanism.

Here, we propose a scenario to account for the crossover between the two dissipative regimes. Theoretical calculations suggest that the vortex cores associated with the $d_{x^2-y^2}$ orbital are considerably larger than those associated with the $d_{z^2}$ orbital~\cite{2023HuangPhysRevB}. In the higher-temperature regime near $T_c(H)$, the larger vortex cores of the $d_{x^2-y^2}$ component may overlap, leading to a strong suppression of its superconducting order parameter, while the $d_{z^2}$ component remains relatively robust between neighboring vortex cores. 
Upon cooling, the vortex cores gradually shrink, allowing the suppressed $d_{x^2-y^2}$ component to recover over a larger spatial region. This evolution of the two orbital components may modify the quasiparticle contribution to vortex-motion dissipation, thereby accounting for the observed crossover between the two dissipative regimes.
\subsection*{C. Vortex glass transition and vortex phase diagram of La$_2$SmNi$_2$O$_{7-\delta}$}
\begin{figure}[b]
\centering
\hspace*{-0.0in}
\includegraphics[width=8.5cm]{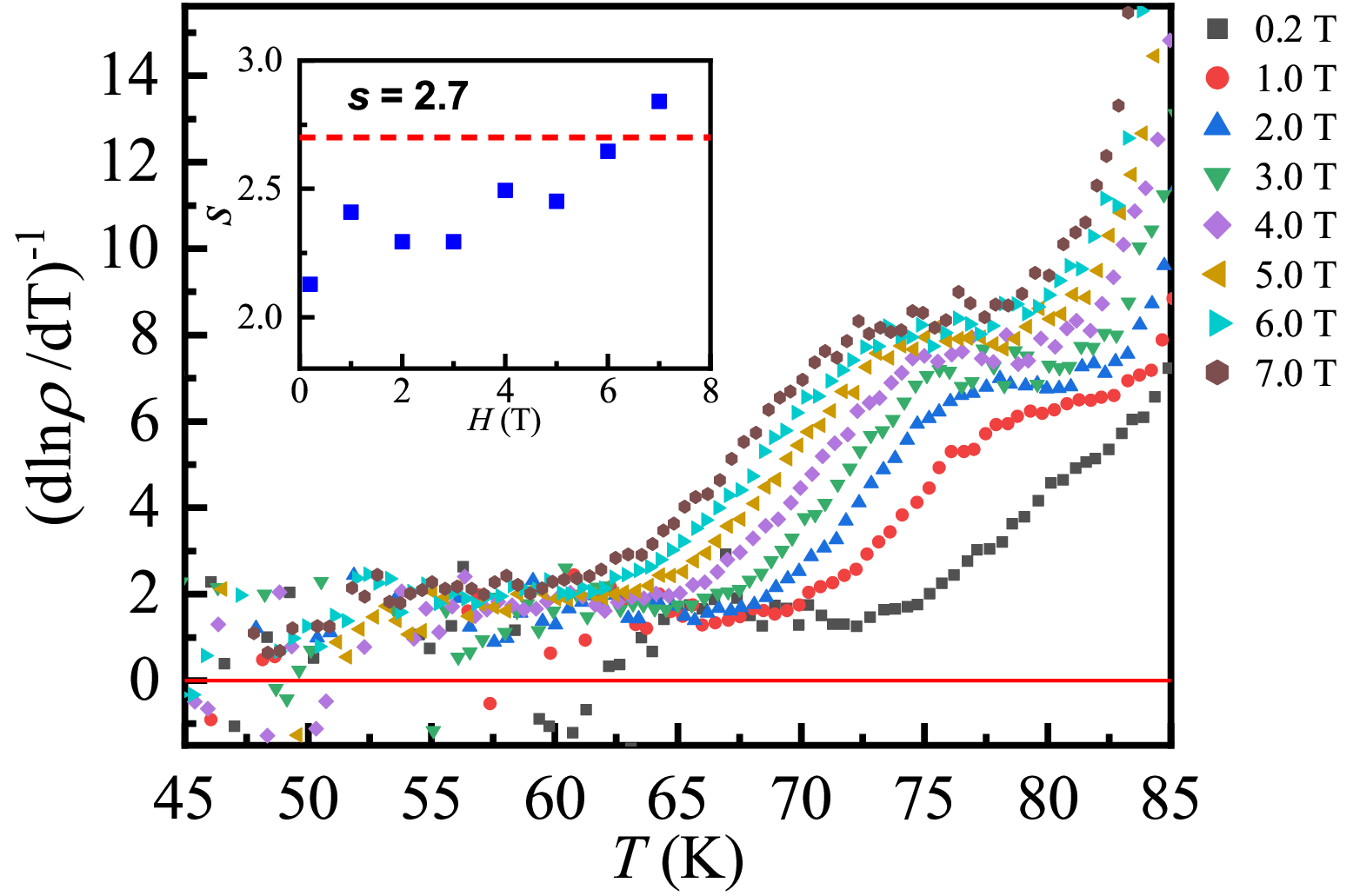}
\caption{
Temperature dependence of $(d\ln\rho/dT)^{-1}$ under magnetic fields up to 7~T applied along the $c$ axis. The inset shows the magnetic-field dependence of the vortex glass critical exponent $s$, where the red dashed line marks $s=2.7$, the reference value for the crossover between 2D and 3D vortex glass behavior.
}
\label{fig:drho_dT}
\end{figure}

To investigate the vortex-glass transition at lower temperatures, we analyze the resistivity using the vortex-glass scaling relation. In the critical region above the glass transition temperature $T_g$, the resistivity is expected to follow a power-law dependence~\cite{1989KochPRL,1989FisherPRL,1991FisherPRB},
\begin{equation}
\rho(T,H)=
\rho_0
\left|
\frac{T}{T_g}-1
\right|^s,
\label{eq:VG}
\end{equation}
where $\rho_0$ is a characteristic resistivity, $s=\nu(z-1)$, and $\nu$ and $z$ are the static and dynamic critical exponents, respectively.
Taking the inverse logarithmic derivative gives
\begin{equation}
\left(
\frac{d\ln\rho}{dT}
\right)^{-1}
=
\frac{T-T_g}{s}.
\label{eq:VGderivative}
\end{equation}
Thus, $(d\ln\rho/dT)^{-1}$ is expected to be linear in temperature in the vortex-glass critical regime. The temperature-axis intercept determines $T_g$, while the inverse slope gives the critical exponent $s$.
Figure~\ref{fig:drho_dT} presents the derivative curves under magnetic fields. Despite noticeable noise near the limit of zero resistance, a roughly linear region can be identified for each field. The extracted exponent $s$, shown in the inset of Figure~\ref{fig:drho_dT}, increases nearly monotonically with magnetic field, ranging between 2.1 and 2.9. 
Notably, the value $s \approx 2.7$ is commonly regarded as the boundary between 2D and 3D vortex glass behavior~\cite{2015VenkateshwarluSST,2017ChioSR}. Therefore, this field dependence suggests a dimensional crossover of the vortex structure, from a more layered character at low fields to increasingly 3D correlations at higher fields.\par

\begin{figure}[t]
\centering
\hspace*{-0in}
\includegraphics[width=9cm]{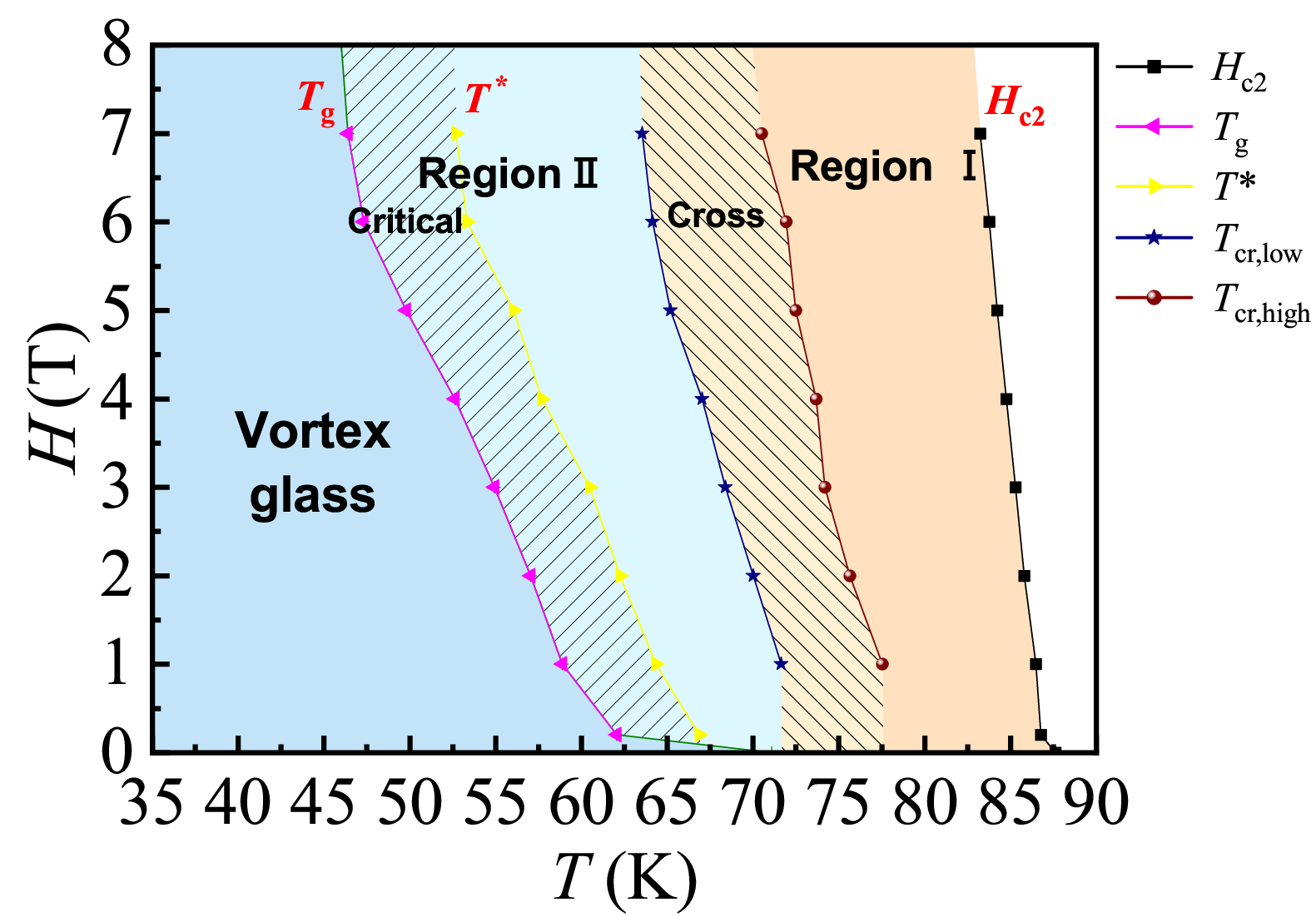}
\caption{
Vortex phase diagram of the La$_2$SmNi$_2$O$_{7-\delta}$ single crystal at 21.2 GPa with $H\parallel c$, determined from magnetotransport measurements. $T_{\mathrm{cr,high}}$ and $T_{\mathrm{cr,low}}$ mark the upper and lower boundaries of the crossover region between the two dissipative regimes, respectively.
}
\label{fig:Phase_diagram}
\end{figure}

Using the characteristic temperatures obtained from the transport analyses, we construct the $H$--$T$ phase diagram of La$_2$SmNi$_2$O$_{7-\delta}$ under pressure, as shown in Figure~\ref{fig:Phase_diagram}. 
The upper critical field $H_{c2}(T)$ is estimated using the criterion of $90\%\rho_n$. 
Below $H_{c2}(T)$, the vortex-liquid regime is divided into three regions: region I, a crossover region, and region II. In this picture, region I is associated with overlapping vortex cores of the $d_{x^2-y^2}$ component, whereas region II corresponds to spatially separated vortex cores, where the resistive dissipation can be described by the TAFF model.
Further cooling drives the vortex system through the glass critical regime and finally into the vortex glass state. The line $T_g(H)$ represents the glass transition temperature, whereas $T^{*}(H)$ denotes the upper boundary of the vortex glass critical regime.
Below $T_g$, the linear resistivity vanishes and the vortices freeze, leading to a zero-resistance state.
\section{Conclusion}
In conclusion, we have investigated the superconducting fluctuations and vortex dynamics of La$_2$SmNi$_2$O$_{7-\delta}$ under pressure. Neither the 2D nor the 3D Ullah--Dorsey scaling provides a satisfactory data collapse near $T_c(H)$, whereas modifying the field exponents in both scaling variables significantly improves the collapse. This anomalous response may originate from the multiorbital nature of the composite superconducting order parameter.
The two-step decrease in resistivity reveals two distinct dissipative regimes, with the low-temperature behavior well described by the conventional TAFF model, while the high-temperature dissipation cannot be accounted for by conventional vortex dynamics.
This crossover may be attributed to the distinct evolution of vortex cores associated with the $d_{x^2-y^2}$ and $d_{z^2}$ orbitals upon cooling. 
Our findings suggest that the multiorbital structure of the superconducting order parameter plays an important role in both critical fluctuations and vortex-motion dissipation in bilayer nickelates.
\section*{ACKNOWLEDGMENTS}
We are grateful for the useful discussions with Ilay Eremin and Andrey Chubukov. This work was supported by the National Key Research and Development Program of China (No. 2022YFA1403201), National Natural Science Foundation of China (Nos. 12434004, 11927809). Junjie Zhang was supported by the Natural Science Foundation of Shandong Province, China (Grant No. ZR2026ZD32) and the National Natural Science Foundation of China (Grants Nos. 12374457 and 12074219). 

\bibliography{reference_LaSm327}

\end{document}